# Anisotropic three-dimensional magnetism in $CaFe_2As_2$

R. J. McQueeney[1,2], S. O. Diallo[2], V. P. Antropov[2], G. D. Samolyuk[2], C. Broholm[3], N. Ni[1,2], S. Nandi[1,2], M. Yethiraj[4], J. L. Zarestky[2], J. J. Pulikkotil[2], A. Kreyssig[1,2], M. D. Lumsden[5], B. N. Harmon[1,2], P. C. Canfield[1,2] and A. I. Goldman[1,2]

[1]Dept. of Physics & Astronomy, Iowa State University, Ames, IA 50011 USA
[2]Ames Laboratory, Ames, IA 50011 USA
[3]Dept. of Physics & Astronomy, Johns Hopkins University, Baltimore, MD 21218 USA
[4]Bragg Institute, ANSTO, PMB 1, Menai NSW 2234, Australia
[5]Oak Ridge National Laboratory, Oak Ridge, TN 37831 USA

## ABSTRACT

Inelastic neutron scattering measurements on the low energy spin waves in $CaFe_2As_2$ show that the magnetic exchange interactions in the Fe layers are exceptionally large and similar to the cuprates. However, the exchange between layers is ~10% of the coupling in the layers and the magnetism is more appropriately categorized as anisotropic three-dimensional, in contrast to the two-dimensional cuprates. Band structure calculations of the spin dynamics and magnetic exchange interactions are in good agreement with the experimental data.

Since the discovery of the new class of the $R$FeAsO [1] and $A$Fe$_2$As$_2$ [2] based superconducting compounds with transition temperatures ($T_c$) up to 55 K [3], there has been an astounding amount of activity intent on characterizing and understanding these materials and the nature of the superconductivity (SC). The observations of antiferromagnetic (AF) ordering,[4] it's subsequent suppression upon chemical doping or changes in stoichiometry,[5] and the eventual appearance of SC [6] are reminiscent of the phase diagram of the high-temperature cuprates.[7] Such similarities might suggest a common origin for SC. In particular, AF spin fluctuations have been proposed as a possible pairing mechanism in both the cuprates [8] and the iron-arsenides.[9] Thus, it is important to compare the details of the magnetic interactions in these two systems. In the cuprates, strong superexchange interactions are present in the AF insulating (parent) phase and give rise to a high AF ordering temperature ($T_N \sim 300$ K).[10] Furthermore, the strength of the magnetic interactions in the cuprates have a strong two-dimensional (2D) anisotropy due to the weak coupling between the CuO$_2$ layers.[11] The iron-arsenide superconductors also have layered structures and high Néel temperatures (100 – 200 K). However, the parent phases of the iron-arsenides are not insulators. Rather, they are metallic and, for the $A$Fe$_2$As$_2$ compounds, the AF ordering is strongly coupled to a structural transition from a high-temperature tetragonal structure to a low temperature orthorhombic structure.[12] One other notable difference between the cuprates and iron arsenides concerns the conditions necessary for SC. While doping charge carriers does indeed suppress AF and lead to superconductivity in both systems, it has recently been shown that pressure alone can destroy the AF state in CaFe$_2$As$_2$ and lead to SC.[13, 14]

Despite these differences, the energy scale and dimensionality (or anisotropy) of the magnetic interactions may actually be quite similar, possibly leading to a common origin for SC in these two families of compounds. In order to move beyond qualitative comparisons and address the relevance of magnetic interactions to SC in the iron-arsenides, direct measurements of the energy scale and anisotropy of the magnetic interactions are necessary. Here we report results from inelastic neutron scattering from $CaFe_2As_2$, both below and above the AF ordering temperature, and demonstrate that the magnetic exchange interactions are exceptionally large, with a similar energy scale as the cuprates. Although the magnetic exchange between the Fe layers is relatively small (> ~10% of the in-plane exchange), it is substantially larger than that found for the cuprates (~0.001%). This anisotropic 3D magnetism is supported by theoretical calculations of the spin dynamics. Despite the first-order magnetostructural transition observed in $CaFe_2As_2$, spin correlations are observed to persist above the AF ordering temperature, attesting to the strength of the magnetism and supportive of a model of frustrated magnetism in the high-temperature tetragonal phase.

$CaFe_2As_2$ is a non-superconducting parent compound that becomes superconducting by either doping [15] or the application of pressure.[14] $CaFe_2As_2$ orders into a columnar-type AF structure (as shown in Fig 1a)) with a simultaneous structural transition from a tetragonal (*I4/mmm*) to an orthorhombic (*Fmmm*) crystal structure below $T_s$ = 172 K with $a$ = 5.51 Å, $b$ = 5.45 Å, and $c$ = 11.66 Å.[12] For the inelastic neutron scattering study, single crystals of $CaFe_2As_2$ were grown out of Sn flux using conventional high temperature solution growth techniques described previously.[16] Crystals were etched in concentrated hydrochloric acid to remove Sn flux from the surface. Measurements were performed on a co-aligned single-crystal composite (~ 1 gram) mounted in the [$H0L$] scattering plane (in orthorhombic notation) with a measured mosaic of 1.5° (full-width-at-half-maximum) for both ($H$00) and (00$L$) reflections. The sample was mounted on the cold finger of a closed-cycle Helium refrigerator. Measurements were performed on the HB-1A and HB-3 triple-axis spectrometers at the High Flux Isotope Reactor at Oak Ridge National Laboratory.

At temperatures well below $T_s$ ($T$ = 14 K), collective spin wave excitations were observed using the HB-3 spectrometer. For the HB-3 measurements, pyrolitic graphite monochromator and analyzer crystals with a fixed final neutron energy $E_f$ = 14.7 meV and horizontal collimations of 48'-40'-80'-120' were used. Figure 1b) shows a schematic drawing of the dispersion surface in the [$H0L$] plane near the magnetic Bragg peak. As discussed below, the energy of the observed excitations is well below the maximum (Brillouin zone boundary) spin wave energy. Therefore, the AF spin waves can be interpreted using the small-**q** approximation with an anisotropic conical dispersion in the [$H0L$] plane and a small energy gap, $\Delta$, as illustrated in Fig. 1b). The gap at the magnetic Bragg peak ($\mathbf{Q}_{AF}$) is clearly demonstrated in Fig. 1c), which displays an energy scan at $\mathbf{Q}_{AF}$ = (103). The background was removed by subtracting the identical scan taken at (1.15,0,3). The data indicate an onset of magnetic scattering above approximately 5-6 meV. Further, Fig. 1c) shows that significant spectral weight is found up to 25 meV (the kinematic limit of the measurement) with little intensity loss, implying that the magnetic energy scale actually extends to much higher energies.

In order to map the dispersion in the vicinity of $\mathbf{Q}_{AF}$, constant-energy scans along the [$H$00] and [00$L$] directions were performed near the $\mathbf{Q}_{AF} = (103)$ and are shown in Figs. 2a) and b). In principle, the constant energy scans should eventually reveal a peak splitting due to dispersion as the energy is increased. However, in CaFe$_2$As$_2$, the energy scale is large enough (with a correspondingly steep dispersion) that the present measurements are unable to observe the splitting up to 25 meV due to the finite instrumental resolution (see fig 1b). Nevertheless, the dispersion does broaden the peaks beyond the spectrometer resolution and the spin wave velocity (the slope of the dispersion in the linear range) can be obtained by fitting the data to a model function convoluted with the resolution function of the instrument. The data was modeled with an anisotropic conical dispersion with an energy gap ($\Delta$) and different spin wave velocities in the Fe layers ($ab$-plane) along $H$ ($v_\parallel$) [17] and perpendicular to the layers along $L$ ($v_\perp$).

$$\hbar\omega(\mathbf{q}) = \sqrt{\Delta^2 + v_\parallel^2(q_x^2 + q_y^2) + v_\perp^2 q_z^2} \qquad (1)$$

Perhaps of greatest importance here is the anisotropy of the velocity ($v_\perp/v_\parallel$), which is a measure of the anisotropy of the magnetic interactions, with $v_\perp/v_\parallel = 1$ representing an isotropic 3D AF and $v_\perp/v_\parallel \approx 0$ a 2D AF. Neutron intensities were represented by a simple Lorentzian response with energy width $\Gamma$ and $1/E$ intensity scaling expected at low temperature.

$$S(\mathbf{Q},E) = \frac{1}{E}\frac{A\Gamma}{(E - \hbar\omega(\mathbf{q}))^2 + \Gamma^2}\delta(\mathbf{Q} - \mathbf{q} - \mathbf{Q}_{AF}) \qquad (2)$$

Here **Q** is the momentum transfer, **q** is the wavevector of the spin wave in the first Brillouin zone, and $A$ is an overall scale factor. The model $S(\mathbf{Q},E)$ was convoluted with the resolution function of the spectrometer and fit to the data using the RESLIB program [18] and the instrument parameters listed above. A small damping factor, $\Gamma$, was fixed at 1 meV (below the resolution limit) to facilitate the numerical convolutions.

The energy gap, $\Delta$, was determined from a fit to the energy scan in fig. 1c) and resulted in a value of $\Delta = 6.9 \pm 0.2$ meV at $T = 14$ K. The in-plane spin wave velocity was fit using the constant energy scans along the [H00] direction (fig. 2a)). The highest energy cuts, at 15, 20, and 25 meV, yield an average in-plane velocity of $v_\parallel = 420 \pm 70$ meV Å. The data at 10 meV are resolution limited and only a lower bound, $v_\parallel > 300$ meV Å, can be determined. The out-of-plane velocity was determined using the data taken along [00$L$] (fig 2b)). The [00$L$] constant energy scans around (103) were, however, obtained in a de-focused resolution condition, and only the 25 meV scan provided a refinable value of $v_\perp = 270 \pm 100$ meV Å. The [00$L$] scans below 25 meV are resolution limited at this value of $v_\perp$, but can be used to determine a minimum velocity parameter, $v_\perp > 200$ meV Å, below which the convoluted model functions are noticeably broader than the data. This minimum velocity condition is consistent with the value obtained at 25 meV. The corresponding dispersion curves are shown in Figs. 2c) and d). The scale factor ($A$) refined to a constant value within 15% for all fits, indicating that the expected $1/E$ dependence of the cross-section is reasonable. As described above, the ratio of $v_\perp/v_\parallel > 0.5$ indicates the 3D nature of magnetism in CaFe$_2$As$_2$, in contrast to the quasi-2D cuprates (where $v_\perp/v_\parallel \approx 0$).

The two principal results of the measurements so far are: (i) that the magnetic exchange interactions in $CaFe_2As_2$ are large and lead to an extremely steep dispersion and; (ii) the ratio of the in-plane and out-of-plane spin wave velocities argue strongly for 3D magnetic interactions in this compound. However, one of the important unanswered questions of the iron arsenic superconductors is the nature of the magnetism above the structural/magnetic transition in the tetragonal phase. The observed Pauli-like paramagnetic susceptibility [16] could indicate reduced or quenched magnetic moments in the tetragonal phase or the response from strongly exchange coupled, but frustrated spins.

To investigate this issue, temperature dependent measurements were made on the HB-1A spectrometer using pyrolitic graphite monochromator and analyzer crystals with a fixed incident neutron energy $E_i$ = 14.7 meV and horizontal collimations of 48'-40'-40'-136'. Upon warming the sample through the transition ($T_S$ = 172 K), Fig. 3a) shows that the gap collapses and spin wave scattering is replaced by broad quasi-elastic scattering. Energy scans were performed at $Q_{AF}$ = (101) and also at (1.2,0,1) for an estimate of the background. The background-subtracted data was corrected for resolution volume and analyzer reflectivity, reduced to the magnetic susceptibility function, and fit to a Lorentzian response, as shown by the solid lines in Fig 3a).

$$\frac{\chi''(\mathbf{Q},E)}{E} = \frac{S(\mathbf{Q},E)\left(1-e^{-E/kT}\right)}{E} = \frac{I(\mathbf{Q})\Gamma}{E^2+\Gamma^2} \qquad (3)$$

Figs 3b) and c), show constant-$E$ scans (at -3 meV) along the [$H$00] and [00$L$] directions as a function of temperature. Spin wave excitations are clearly observed at 140K and are replaced, above the transition, by correlated spin fluctuations. Intensity persists near $\mathbf{Q}_{AF}$, but the peaks are much broader above the transition indicating a short correlation length (~ 30 Å). These data demonstrate that, despite the strong first-order nature of the magnetostructural transition,[12] the large exchange coupling gives rise to short-ranged, but dispersive (paramagnon) excitations that exist up to ~200 K. At temperatures above ~200 K, the scattering becomes more smeared out along the $H$ and $L$ directions indicating that the fluctuations are becoming uncorrelated.

We now turn to theoretical calculations of the spin dynamics in CaFe$_2$As$_2$. There are several papers that address the nature of the magnetic interactions in the iron arsenic superconductors.[19, 20] Despite the metallic nature, the interactions are often discussed in a localized 2D Heisenberg picture where the interlayer exchange is assumed to be negligible. Based on the disagreement of our experimental results with this notion, the exchange couplings were calculated using Green's function formalism within the Atomic Sphere Approximation.[21] Using the experimental structure parameters,[12] these calculations predict the observed columnar AF structure with a moment size of $gS = 1.33$ $\mu_B$ with $g \approx 2$ (somewhat larger than the observed moment of 0.8 $\mu_B$). The calculated magnetic interactions are long-ranged, as one might expect for metallic CaFe$_2$As$_2$. 'Frozen' magnon calculations [21, 22] of the spin wave velocities ($v_\parallel = 390$ meV Å, $v_\perp = 190$ meV Å) are in agreement with the observed values.

The theoretical calculations can also be discussed within the context of a Heisenberg model consisting of only nearest-neighbor (NN) and next-nearest-neighbor (NNN) interactions. The calculated NN couplings in the three orthorhombic directions are AF and anisotropic,[23] with $SJ_{1a} = 41$ meV, $SJ_{1b} = 10$ meV, and $SJ_{1c} = 3$ meV (see fig. 1a)). Appreciable NNN exchange ($SJ_2 = 21$ meV) is sufficiently large to stabilize the columnar AF structure given that $(J_{1a} + J_{1b})/2 : J_2 = 1.2$ (fulfilling the condition that $J_1/J_2 < 2$ for the tetragonal structure[20]). The rather significant exchange along the $c$-axis confirms the 3D nature of the system and indicates that $J_c/J_{1a} \sim 10\%$ for CaFe$_2$As$_2$. This is sufficiently closer to 3D than the cuprates, where $J_c/J_{1a} \sim 10^{-5}$. For the Heisenberg model, the calculated spin velocities are $v_{\parallel} = aS(J_{1a} + 2J_2)\sqrt{1 + J_{1c}/(J_{1a} + 2J_2)} = 450$ meV Å and $v_{\perp} = cSJ_{1c}\sqrt{1 + (J_a + 2J_2)/J_{1c}} = 190$ meV Å and the calculated velocity ratio is

$$\frac{v_{\perp}}{v_{\parallel}} = \frac{c}{a}\sqrt{\frac{J_{1c}}{J_{1a} + 2J_2}} \approx 0.4 \qquad (4)$$

Velocities within the NNN Heisenberg model are in surprisingly close agreement with the more general frozen magnon calculations, despite the neglect of more distant neighbors. Nonetheless, all of the calculated quantities are in agreement with the data and support anisotropic 3D magnetism in CaFe$_2$As$_2$. Neutron measurements on SrFe$_2$As$_2$ also report significant inter-layer magnetic interactions.[24] However, the interlayer coupling in CaFe$_2$As$_2$ is several times larger then that calculated in the closely related FeSe based superconductors,[25] indicating that the degree of the interlayer coupling may be system dependent.

Finally, the calculations predict several details about the magnetic Hamiltonian that are presently beyond experimental confirmation. Using a numerical "torque" technique,[22] a 2.1 meV spin gap is estimated from the spin-orbit coupling. This is significantly smaller than the measured gap (~7 meV) and indicates the presence of additional spin space anisotropies. The theoretical calculations also characterize the "itinerancy" of the magnetic interactions in this system as being similar to face-centered cubic Fe (with an adiabatic parameter of 0.2 [26]). Thus we can classify $CaFe_2As_2$ as marginally itinerant and at the borderline between itinerant and localized behavior. In this limit, the calculations of the magnetic susceptibility indicate substantial longitudinal magnetic fluctuations similar to the FeSe system.[25]

In conclusion, despite the similarity of the magnetic energy scale in $CaFe_2As_2$ and the cuprates, there are substantial differences in the magnetic Hamiltonians. In the cuprates, the 2D Heisenberg model is dominant, and only lifted by small X-Y anisotropy and antisymmetric exchange interactions.[27] In the $CaFe_2As_2$, there is a substantial interlayer coupling and the presence of a spin gap leads to additional weaker terms related to spin space anisotropies. Strictly 2D magnetic systems are characterized by large quantum spin fluctuations, and such spin fluctuations have been proposed as a possible mechanism for superconductivity. The 3D magnetic interactions in $CaFe_2As_2$ should suppress the quantum fluctuations. However, the large exchange interactions combined with frustration may be sufficient to support spin fluctuations in superconducting compositions.

**Acknowledgments** The authors acknowledge useful discussions with J. Schmalian, D. Johnston, and A. Christianson. Work is supported by the U. S. Department of Energy Office of Science under the following contracts: Ames Laboratory under Contract No. DE-AC02-07CH11358, and Oak Ridge National Laboratory, which is managed by UT-Batelle LLC, under Contract No. DEAC00OR22725.

**Figure captions**

FIG. 1. (a) The magnetic structure (showing Fe atoms only) and exchange interactions in CaFe$_2$As$_2$. (b) Schematic picture of the antiferromagnetic conical spin wave dispersion surface. The shaded plane indicates a constant energy surface, and the shaded ellipses represent the orientation of the resolution ellipsoid for H and L scans. (c) Intensity as a function of neutron energy loss at the magnetic zone center, $Q_{AF}$ = (103). The solid black line shows fits to the data using the model dispersion (eq. (1)) convoluted with the instrumental resolution function.

FIG. 2. (color online) (a) Intensity along the *H*-direction near $Q_{AF}$ = (103) at several different energies. Scans at different energies are vertically offset by 160 counts.
(b) Constant energy scans along *L*. In (a) and (b), the solid black lines are fits to the data using the model dispersion (eq. (1)) convoluted with the instrumental resolution function. Dashed lines are estimates of the resolution lineshape at 25 meV. (c) The plotted model dispersion along the [*H*00] direction (black line). (d) The model dispersion along the [00*L*] direction is plotted as a shaded region, indicating that the data gives only a lower bound to the spin wave velocity along *L*. In (c) and (d), fitted data points are indicated by black circles and lower bounds to the dispersion are indicated by vertical lines connected by a bar.

FIG. 3. (color online) (a) The dynamic magnetic susceptibility of CaFe$_2$As$_2$ as a function of energy at $Q_{AF}$ = (101) for several temperatures. The 173 K data are offset vertically by 0.06 units and higher *T* data by 0.03 units. Data taken above the transition at $T_s$ = 172

K were fit (solid black line) to a Lorenztian response function plus a Gaussian function (for the elastic scattering peak at $E = 0$ meV). The dashed line at 140 K is a guide to the eye. (b) Scans along the [$H$01] direction at $E = -3$ meV and at temperatures below the transition (red, 140 K), just above the transition (green, 178 K), and well above the transition (blue, 200 K). Solid black lines show fits to a Gaussian peak shape plus background function. The temperature independent peak at $H = 0.8$ has unknown origin and is treated as background. (c) Scans along the [10$L$] direction at $E = -3$ meV and at the same temperatures as (b).

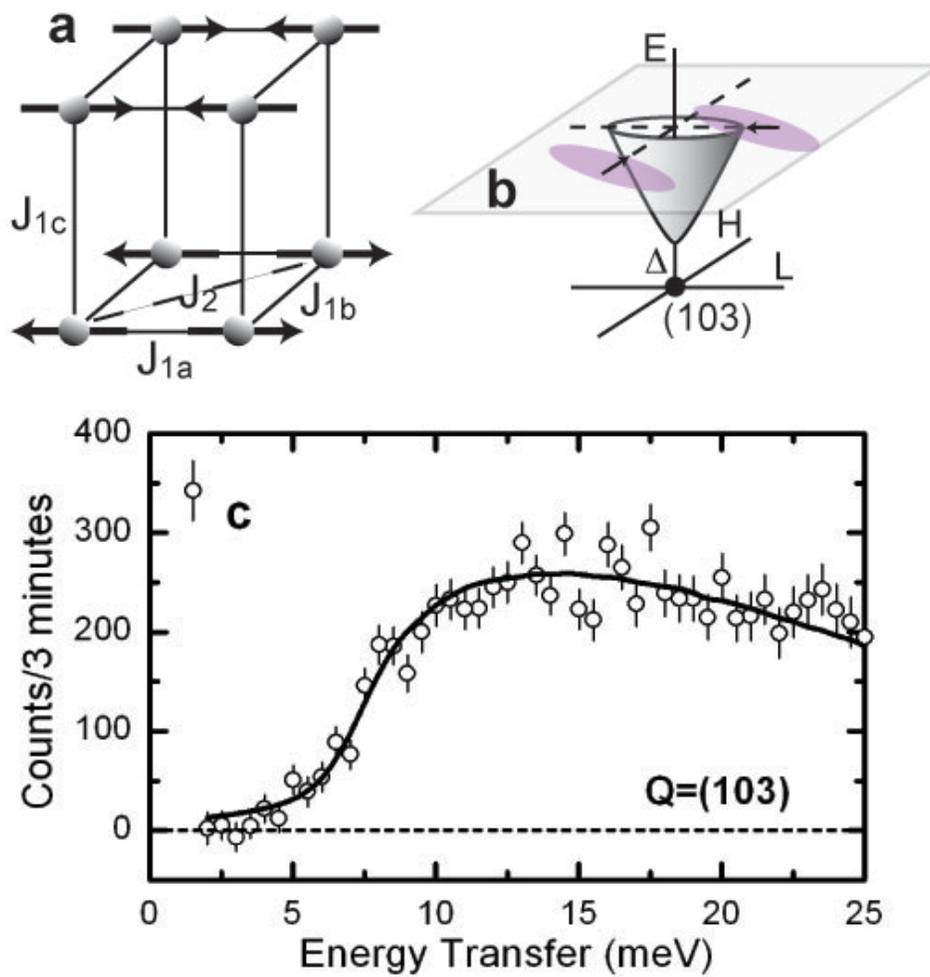

FIG. 1

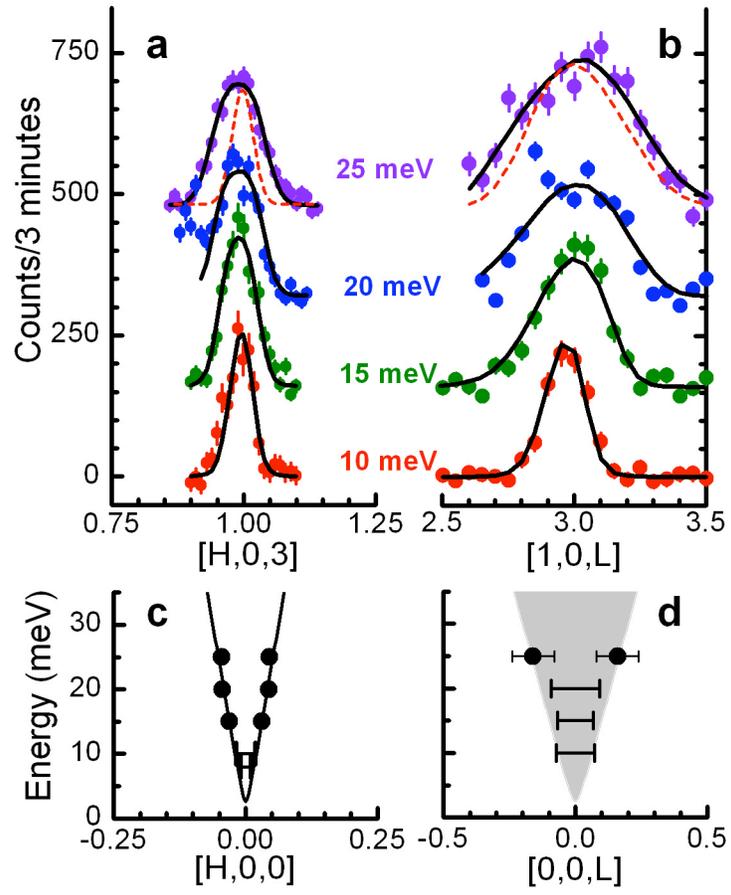

FIG 2

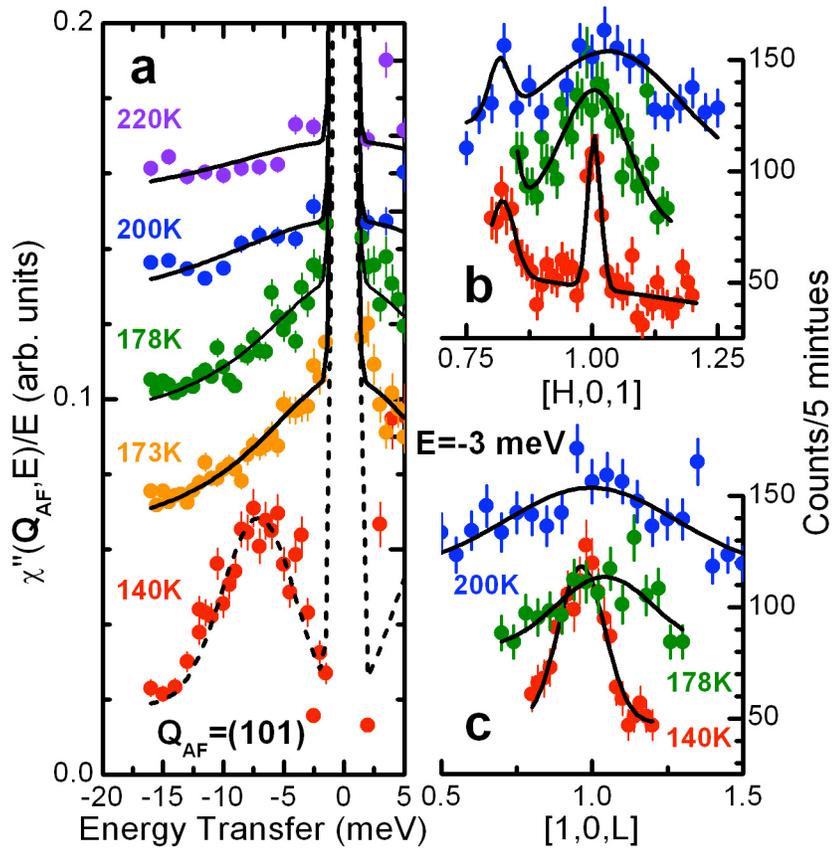

FIG 3